\documentclass{llncs} 
\usepackage[latin1]{inputenc}
\usepackage[english]{babel}
\usepackage{mathrsfs}

\usepackage{amsmath,amssymb,amstext,amsgen,amsbsy,amsopn,
 epsfig,graphicx,bbm}

\newtheorem{thm}{\bf{Theorem}}
\newtheorem{defi}[thm]{\bf{Definition}}
\newtheorem{lemme}[thm]{\bf{Lemma}}
\newtheorem{prop}[thm]{\bf{Proposition}}
\newtheorem{exm}[thm]{\bf{Example}}
\newtheorem{coco}[thm]{\bf{Corollary}}

\newenvironment{demo}{ \textbf{Proof:} }{$\blacksquare$ }

\newcommand{\debco}{\begin{coco} }
\newcommand{\debarr}{\begin{array} }
\newcommand{\debctr}{\begin{center} }
\newcommand{\debdef}{\begin{defi} }
\newcommand{\deblem}{\begin{lemme} }
\newcommand{\debpro}{\begin{prop} }
\newcommand{\debthm}{\begin{thm} }
\newcommand{\debexm}{\begin{exm} }
\newcommand{\debdem}{\begin{demo} }
\newcommand{\debeqn}{\begin{eqnarray}}  
\newcommand{\debeqno}{\begin{eqnarray*}}

\newcommand{\finco}{\end{coco} }
\newcommand{\finarr}{\end{array} }
\newcommand{\finctr}{\end{center} }
\newcommand{\findef}{\end{defi} }
\newcommand{\finlem}{\end{lemme} }
\newcommand{\finpro}{\end{prop} }
\newcommand{\finthm}{\end{thm} }
\newcommand{\finexm}{\end{exm} }
\newcommand{\findem}{\end{demo} }
\newcommand{\fineqn}{\end{eqnarray}}
\newcommand{\fineqno}{\end{eqnarray*}}

\newcommand{\debitem}{\begin{itemize} }
\newcommand{\finitem}{\end{itemize} }

\newcommand{\inc}{\subseteq}
\newcommand{\ppt}{\leqslant}
\newcommand{\pgd}{\geqslant}

\newcommand{\Equ}{ \Longleftrightarrow}

\newcommand{\EXPSPACE}{{\sc expspace}}
\newcommand{\EXPTIME}{{\sc exptime}}

\newcommand{\gr}{ \mathscr{G}}

\newcommand{\sr}{ \mathscr{S}}

\newcommand{\eps}{ \varepsilon}
\newcommand{\ie}{{\it i.e.}}

\newcommand{\ti}{\times}

\newcommand{\zero}{Player~0}
\newcommand{\one}{Player~1}

\newcommand{\g}{\gamma}

\newcommand{\m}{\mu}

\renewcommand{\r}{\rho}
\newcommand{\s}{\sigma}

\renewcommand{\t}{\tau}

\newcommand{\D}{\Delta}

\newcommand{\G}{\Gamma}

\newcommand{\T}{\theta}

\renewcommand{\S}{\Sigma}

\newcommand{\Oo}{{\cal O}}

\newcommand{\first}{\mathit{first}}
\newcommand{\last}{\mathit{last}}
\newcommand{\counter}{\mathit{counter}}
\newcommand{\equal}{\mathit{equal}}
\newcommand{\Succ}{\mathit{succ}}
\newcommand{\struct}[1]{\langle #1 \rangle}

\title{The Complexity of Games on Higher Order Pushdown Automata
\thanks{This research has been partially supported
   by the European Community Research Training Network
   ``Games and Automata for Synthesis and Validation'' (GAMES),
   (contract HPRN-CT-2002-00283), see \tt{www.games.rwth-aachen.de}. } }
\author{Thierry Cachat\inst{1} \and Igor Walukiewicz\inst{2}} 
\institute{LIAFA/CNRS UMR 7089 \& Universit{\'e} Paris 7, France
	\\ \texttt{Firstname.Lastname@liafa.jussieu.fr}
\and
LaBRI, Université Bordeaux-1, France
\email{igw@labri.fr}
}

\begin{document}

\maketitle

\pagestyle{plain}  

\begin{abstract}
  We prove an $n$-\EXPTIME\ lower bound for the problem of deciding
  the winner in a reachability game on Higher Order Pushdown Automata
  (HPDA) of level $n$. This bound matches the known upper bound for
  parity games on HPDA. As a consequence the $\m$-calculus model
  checking over graphs given by $n$-HPDA is $n$-\EXPTIME\ complete.
\end{abstract}

%
%
%
\section{Introduction} 
%
%
%

Higher Order Pushdown Automaton (HPDA) is a classical model of
computation \cite{engel,engel2} that has recently regained attention.
In \cite{teo} it has been proved that the MSO theory of the
computation trees of HPDA is decidable. Then in \cite{hierar} a new 
family of infinite graphs, also with a decidable MSO theory, 
has been introduced, which is closely related to HPDA (see
\cite{icalp03,arnaudS}). See also other approaches in 
\cite{MeyerBouajjani04,Carayol05}. Up to now the Caucal hierarchy
of \cite{hierar} is essentially the largest class of graphs with a decidable 
MSO theory. But these decidability results have non-elementary complexity,
even for a fixed level of the hierarchy. Considering $\mu$-calculus
model-checking and parity games allows to have better complexity bounds.

We consider the question of deciding a winner in a reachability game
given by a HPDA. It was shown by the first author~\cite{icalp03} that
parity games on $n$-HPDA's can be solved in $n$-\EXPTIME. This also
gives $n$-\EXPTIME\ algorithm for the $\m$-calculus model checking over
such graphs. Here we complement the picture by showing that even
reachability games are $n$-\EXPTIME\ hard on $n$-HPDA's, thereby showing
$n$-\EXPTIME\ completeness for game solving and $\m$-calculus model
checking over $n$-HPDA's.

It was already shown by the second author in~\cite{igor} that pushdown
games (on 1-HPDA) are \EXPTIME-complete. We extend the technique
with codding big counters, following the notation from~\cite{ltrl},
where the computation of space bounded Turing machines are written
with the help of 1-counters of $n$-bits, $2$-counters of $2^n$ bits
and so on. The expressive power of HPDA is used to ``copy'' parts of
the store and check equality of big counters.

In the next section we present the definitions of game and HPDA.
In Section~\ref{sec:Engel} we prove the lower bound using a reduction
from the word problem for alternating HPDA and a result by Engelfriet.
The rest of the paper is devoted to an alternative, self contained and 
hopefully simple, proof of the lower bound.
Using HPDA we show in Section~\ref{sec:2count} how to 
handle counters of level 1 and 2, and then of higher levels. 
In Section~\ref{sec:TM} we use counters to encode configurations of 
Turing Machines and prove the lower bound.

We assume that the reader is familiar with the basic notions of 
games (see \cite{DAGST} for an overview).

%
%
%
\section{Definitions: Game, HPDS} 
%
%
%

\subsection{Game}\label{sub-parity}  
%
%
An {\em arena} or {\em game graph} is a tuple $(V_0,V_1,E)$, where
$V=V_0\uplus V_1$ is a set of vertices partitioned into vertices of
Player~0 and vertices of Player~1, and $E\inc V\ti V$ is a set of
edges (directed, unlabeled).  Starting in a given initial vertex
$\pi_0\in V$, a play in $(V_0,V_1,E)$ proceeds as follows: if $\pi_0
\in V_0$, Player~0 makes the first move to $\pi_1$ with $\pi_0 E
\pi_1$, else Player~1 does, and so on from the new vertex $\pi_1$. A
play is a (possibly infinite) maximal sequence $\pi_0 \pi_1 \cdots$ of
successive vertices.  For the winning condition we consider
reachability: a subset $F\inc V$ is fixed, and \debeqno \mbox{\zero\ 
  wins }\pi \mbox{ iff }\exists i: \pi_i\in F\ .  \fineqno As soon as
$F$ is reached, the play stops. The play can also stop when a position
is reached with no outgoing edges. In this case the player who is
supposed to move loses.  A {\em strategy} for Player~0 is a function
associating to each prefix $\pi_0 \pi_1 \cdots \pi_n$ of a play such
that $\pi_n\in V_0$ a ``next move'' $\pi_{n+1}$ with $\pi_n E
\pi_{n+1}$. We say that Player~0 wins the game from the initial vertex
$\pi_0$ if he has a {\em winning strategy} for this game: a strategy
such that he wins every play. 

\subsection{Higher Order Pushdown System}
%
%

We recall the definition from \cite{teo} (which is equivalent to the
one from \cite{engel}), where we slightly change the terminology. A
{\em level 1 store} (or {\em 1-store}) over an alphabet $\G$ is an
arbitrary sequence $\g_1\cdots \g_\ell$ of elements of $\G$, with
$\ell\pgd 0$. A {\em level $k$ store} (or {\em $k$-store}), for $k\pgd
2$, is a sequence $[s_1]\cdots [s_\ell]$ of $(k-1)$-stores, where
$\ell\pgd 0$.  The following operations can be performed on $1$-store:
\debeqno push_1^\g(\g_1\cdots \g_{\ell-1} \g_\ell)& := & \g_1 \cdots
\g_{\ell-1}
\g_\ell \g \mbox{ for all }\g\in\G\ ,\\
pop_1(\g_1 \cdots \g_{\ell-1} \g_\ell)& := & \g_1 \cdots \g_{\ell-1}\ ,\\
top(\g_1 \cdots \g_{\ell-1} \g_\ell)& := &\g_\ell\ .  
\fineqno 
If $[s_1]\cdots [s_\ell]$ is a store of level $k>1$, the following 
operations are possible: 
\debeqno
push_k([s_1]\cdots [s_{\ell-1}][s_\ell])& := &[s_1]\cdots[s_{\ell-1}][s_\ell][s_\ell]\ ,\\
push_j([s_1]\cdots [s_{\ell-1}][s_\ell])& :=
&[s_1]\cdots[s_{\ell-1}][push_j(s_\ell)]
\mbox{ if }2\ppt j< k\ ,\\
push_1^\g([s_1]\cdots[s_{\ell-1}][s_\ell])& :=
&[s_1]\cdots[s_{\ell-1}][push_1^\g(s_\ell)]
\mbox{ for all }\g\in\G\ ,\\
pop_k([s_1]\cdots[s_{\ell-1}][s_\ell])& := &[s_1]\cdots[s_{\ell-1}]\ ,\\
pop_j([s_1]\cdots[s_{\ell-1}][s_\ell])& := &
[s_1]\cdots[s_{\ell-1}][pop_j(s_\ell)]\mbox{ if }1\ppt j< k\ ,\\
top([s_1]\cdots[s_{\ell-1}][s_\ell])& := &top(s_\ell)\ .  
\fineqno
The operation $pop_j$ is undefined on a store, whose top store of level
$j$ is empty. Similarly $top$ is undefined on a store, whose top
1-store is empty.
We will consider ``bottom store symbols'' $\bot_j\in\G$ at
each level $1\ppt j\ppt k$. When a $j$-store is empty, implicitly 
its top symbol is $\bot_j$. These symbols can neither be erased nor
``pushed''.
Given $\G$ and $k$, the set $Op_k$ of operations (on a store) 
of level $k$ consists of:
\debeqno
        push_j \mbox{ for all }2\ppt j\ppt k,\ 
        push_1^\g \mbox{ for all }\g\in\G,\ 
        pop_j\mbox{ for all }1\ppt j\ppt k\mbox{, and }
        skip\ .
\fineqno
The operations $push_j$, allowing to ``copy'' a part of the store, 
are responsible for the fact that the hierarchy of HPDS is strict.
A {\em higher order pushdown system} of level $k$ (or { $k$-HPDS}) is a tuple 
$H=(P,\G,\D)$ where $P$ is the finite set of control locations, 
$\G$ the finite store alphabet,
and $\D \inc P\times \G \times P\times Op_k$
the finite set of (unlabeled) transition rules.
We do not consider HPDS as accepting devices, hence there is no input
alphabet. The name HPDS is derived from Pushdown System (PDS), it is a HPDA
with unlabeled transitions.
A {\em configuration} of an $k$-HPDS $H$ is a pair $(p,s)$ where $p\in P$ and
$s$ is an $k$-store. The set of $k$-stores is denoted $\sr_k$.
A HPDS $H=(P,\G,\D)$ defines a {\em transition graph } $(V,E)$,
where $V=\{(p,s): p\in P, s\in\sr_k\}$ is the set
of all configurations, and
\debeqno
        (p,s)E(p',s') \Equ \exists (p,\g,p',\theta)\in\D: 
        top(s)=\g \mbox{ and } s'=\theta(s)\ .
\fineqno
For our constructions it would be simpler to assume that $k$-HPDS can
work also on stores of lower levels, in particular on $1$-stores. Of
course we can always simulate a $j$-store, for $j<k$  with an
$k$-store but in the notation it requires some additional
parenthesis that make it less readable.

To define a game on the graph of a HPDS, we assign a player 
to each control state, and we consider an initial configuration:
a {\em game structure on a HPDS $H$} is a tuple $\gr=(H,P_0,P_1,s_0)$,
where $P=P_0\uplus P_1$ is a partition of the control states of $H$, 
and $s_0\in \sr_k$.
This extends naturally to a partition of the set of configurations: 
with the notations of Section~\ref{sub-parity}, $V_0=P_0\times\sr_k$,
$V_1=P_1\times\sr_k$, and $E$ is defined above.

%
%
%
\section{Reduction from the Word Problem} \label{sec:Engel}
%
%
%
Higher Order Pushdown Automata were originally designed to recognize languages.
In the usual way transitions can be labeled by letters from an input alphabet
$A$. A non-deterministic HPDA is defined like a HPDS above except that
$\D \inc P\times \G\ti (A\cup\{\eps\}) \times P\times Op_n$. A transition
can ``read'' a symbol from the input word or stay on the same place. The edges of
the transition graph are labeled accordingly, and a word is accepted iff there
exist a path from an initial configuration to a final configuration. Here
the initial configuration can be chosen arbitrarily and the final configurations
are defined by the control state.

In an alternating (one-way) HPDA each control state is either existential
(in $P_0$) or universal (in $P_1$). 
A computation is a tree, from which the root is $(p_0,s_0,0)$
where $p_0$ is the initial control state, $s_0$ is the initial store content,
and $0$ represents the leftmost position of the input word. If the input word is 
$w=w_1\dots w_{|w|}$, then every non-leaf node $(p,s,i)$ in the tree 
must satisfy the following.
\debitem
\item If $p\in P_0$ then {\em there is} a transition $(p,\g,a,p',\T)\in \D $
  such that $top(s)=\g$ and 
  \debitem
  \item either $a=w_{i+1}$ and the node $(p,s,i)$ has one child $(p',\T(s),i+1)$, 
  \item or $a=\eps$ and the node $(p,s,i)$ has one child $(p',\T(s),i)$.
  \finitem
\item If $p\in P_1$ then 
  \debitem
  \item {\em for each} transition $(p,\g,a,p',\T)\in \D $ such that 
    $top(s)=\g$ and $a=w_{i+1}$, the node $(p,s,i)$ has a child $(p',\T(s),i+1)$,
  \item and {\em for each} transition $(p,\g,\eps,p',\T)\in \D $ such that 
    $top(s)=\g$, the node $(p,s,i)$ has a child $(p',\T(s),i)$.
  \finitem
\finitem
A word $w$ is accepted if there exists a computation tree such that every
leaf is (labeled by) an accepting state. 

It is well known that there is strong connections between alternation and games
(see e.g. \cite{DAGST}) but these connections depends very much on the context
(finite/infinite words, epsilon-transitions allowed or not, \dots).

Let $Tower$ stand for the ``tower of exponentials''
function, \ie, $Tower(0,n)=n$ and $Tower(k+1,n)=2^{Tower(k,n)}$.
One of the results of \cite{engel2} is that given $k>0$, the class of
languages of alternating level $k$ HPDA is the class
\debeqn
       \bigcup_{d>0} DTIME(Tower(k,d n))  \label{eqn:compl}
\fineqn
where $n$ is the length of the input word.

Given a $k$-HPDA $H=(P,\G,\D)$ and a word $w$, our aim is to define a $k$-HPDS 
$G$ and a game structure on $G$ such that \zero\ wins if and only if $w$
is accepted by $H$. Because in the game there is no input word, the idea is to
encode $w$ in the control states and in the transitions of $G$.
Let $Q=P\ti[0,|w|]$ and $G=(Q,\G,\D')$ where
\debeqno
       \D' & = & 
       \{ ((p,i),\g,(p',i+1),\T): (p,\g,a,p',\T)\in \D \mbox{ and } w_{i+1}=a\}\cup\\
       && \{ ((p,i),\g,(p',i),\T): (p,\g,\eps,p',\T)\in \D \}
\fineqno
The set $Q_0$ of control states where \zero\ moves is $P_0\ti[0,|w|]$, corresponding
to existential states. 
The set $Q_1$ where \one\ moves is $P_1\ti[0,|w|]$, corresponding
to universal states. The goal set $F$ is given by the final state(s) of $H$.
\debpro
      Given an alternating (one-way) HPDA $H$ and an input word $w$ one can construct 
      in polynomial time a game structure on a HPDS of the same level and 
      whose size is linear in $|H|.|w|$.
\finpro
Note that this proposition can be easily extended to alternating two-way HPDA.
From the results of \cite{engel2} (see (\ref{eqn:compl}) above) it
follows that for every $k>0$ and $d>0$ there is a HPDA $H$ of level 
$k$ such that the word problem for $H$ cannot be decided in less than 
$DTIME(Tower(k,d n))$. It follows from this fact and the previous proposition that 
a game on a HPDS $G$ of level $k$ and size $|G|$ cannot be solved in less than
$DTIME(Tower(k,|G|))$.
\debthm
      Reachability games on $k$-HPDS are $k$-\EXPTIME\ hard. \label{thm:hard}
\finthm
Note that given an alternating HPDA $H$, one can simply remove the transition labels 
and the input alphabet, keeping the same set of control states. 
The game structure $G$ obtained is such that: if some word is accepted by $H$ then
the game is won by \zero, but the converse is not true. So there is no clear link
between the emptiness problem and the game problem.
The situation is different if one considers infinite words (a Büchi acceptance 
condition), a unary alphabet and no epsilon-transitions.

%
%
%
\section{Counters} \label{sec:2count}
%
%
%
In the rest of the paper we give an alternative proof of Theorem~\ref{thm:hard}.
Our final aim will be to encode computation of $k$-\EXPSPACE\ bounded
alternating Turing machines using $k$-HPDS. As a preparatory step we
will show that using $k$-HPDS we can manipulate numbers of up to
$Tower(k,n)$.

\subsection{Alphabets} 
%
%
For each index $i\geq 1$ we consider the alphabet
$\Sigma_i=\{a_i,b_i\}$, where $a_i$ and $b_i$ are associated to
$a$ and $b$ when regarded as letters of the Turing machine, and
to $0$ and $1$ when regarded as bits (respectively). This conventions
will be used through-out the rest of the paper. 

\subsection{$2$-counters}
%
%

As an introductory step we will show that we can count up to $2^{2^n}$
using $2$-store.

\debdef
Given $n>0$, a {\em $1$-counter} of length $n$ is a word 
\debeqno
        \sigma_{n-1}\cdots \sigma_1\sigma_0\in (\Sigma_1)^n\ , 
\fineqno
it represents the number $\sum_{i=0}^{n-1}\sigma_i 2^i$ (recall that
the letter $a_1$ represents $0$ and the letter $b_1$ represents $1$.)
\findef
So we use counters of $n$ bits, and the parameter $n$ is now fixed 
for the rest of this section without further mentioning. 
\debdef
A {\em $2$-counter} is a word
\debeqno
        \sigma_k \ell_k \cdots  \sigma_1 \ell_1 \sigma_0 \ell_0\ , 
\fineqno
where $k=2^n-1$, 
 for all $i\in [0,2^n-1]$ we have
$\sigma_i \in \Sigma_2$ and $\ell_i\in (\Sigma_1)^n$ is a 
$1$-counter representing the number $i$.
This $2$-counter represents the number 
$\sum_{i=0}^{2^n-1}\sigma_i 2^i$. 
\findef
We will see how to force \zero\ to write down a proper counter on the
store. More precisely we will define states that we call tests. From
these states it will be possible to play only a finite game which will
be designed to test some properties of the stack. For example, \zero\ 
will win from $(\counter_i,u)$ iff a suffix of $u$ is an $i$-counter.
\\[1 em]
From a configuration $(\counter_1,u)$ we want \zero\ to win iff on the
top of the stack there is a $1$-counter; more precisely when $u$ has
a suffix $\s_2 v\s_2'$ for $v\in (\S_1)^n$ and $\s_2,\s_2'\in\S_2$. To
obtain this we let \one\ pop $n+2$ letters and win if inconsistency is
discovered; if no inconsistency is found then \zero\ wins. Similarly
we can define $\first_1$ and $\last_1$ from which \zero\ wins iff on
the top of the stack there is a $1$-counter representing $0$ and
$2^n-1$ respectively.
\\[1 em] 
In a configuration $(\equal_1,u)$ we want \zero\ to win iff the two
topmost $1$-counters have the same value; more precisely when a suffix
of the stack $u$ is of the form $\s_2v\s'_2v\s''_2$ with $v\in
(\S_1)^n$, $\s_2,\s'_2,\s''_2\in \S_2$. In the state $\equal_1$ \one\ 
has the opportunity either to check that there are no two $1$-counters
on the top of the stack (which is done with $\counter_1$), or to
select a position where he thinks that the counters differ. To do this
he removes from the stack up to $n$ letters in order to reach a
desired position.  The bit value of this position is stored in the
control state and then exactly $n+1$ letters are taken from the stack.
\one\ wins iff the letter on the top of the stack is different from
the stored bit value; otherwise \zero\ is the winner.
\\[1 em] 
Similarly, in a configuration $(\Succ_1,u)$ \zero\ wins iff the two
topmost $1$-counters represent successive numbers; more precisely when
$u$ has a suffix of the form $\s_2v\s'_2v'\s''_2$ with $v,v'\in
(\S_1)^n$ representing consecutive numbers, and $\s_2,\s'_2,\s''_2\in
\S_2$. As before \one\ has an opportunity to check if the stack does
not end with two $1$-counters. The other possibility is that \one\ can
select a position where he thinks that the value is not right.  First
he can ``pop'' any number of letters. During this process, the control
state remembers whether the letter $b_1$ (which represents $1$) has
already been seen: because lowest bits are popped first, as long as
$a_1$ are popped, we know the corresponding letter in the other
counter should be a $b_1$. After the first $b_1$, the letters should
be the same in the other counter. Then exactly $n+1$ letters or popped
(including $\s'_2$) and \one\ wins if the letter is not right;
otherwise \zero\ wins.
\\[1 em] 
Starting from a configuration $(\counter_2,u)$ we want \zero\ to win
iff on the top of the stack there is a $2$-counter; more precisely
when $u$ has a suffix $\s_3v\s'_3$ with $\s_3,\s'_3\in \S_3$ and $v$ a
$2$-counter. A $2$-counter is a sequence of $1$-counters, and the task
of \one\ is to show that $u$ has no suffix of the right form. One way
to do this is to show that $u$ does not end with a $1$-counter or that
this last counter does not have value $2^n-1$. This \one\ can do with
$\last_1$ test.  Otherwise \one\ can decide to show that there is some
part inside the hypothetical $2$-counter that is not right. To do this
he is allowed to take letters from the stack up to some $\S_2$ letter
at which point he can check that the two topmost counters have wrong
values (using test $\Succ_1$). This test can be performed only if
\zero\ does not claim that the counter on the top represents $0$.  If
\zero\ claims this then \one\ can verify by using test $\first_1$.  It
should be clear that if $u$ does not end with a $2$-counter then \one\ 
can make the right choice of a test and win. On the other hand if $u$
indeed ends with a $2$-counter then \zero\ wins no matter what \one\ 
chooses.  Similarly we can define $\first_2$ and $\last_2$ from which
\zero\ wins iff the top of the store is a $2$-counter representing
values $0$ and $2^{2^n}-1$ respectively.
\\[1 em] 
%
%
Next we want to describe $\equal_2$ test for which we will need the
power of $2$-stores. We want \zero\ to win from a configuration
$(\equal_2,u)$ iff there is a suffix of $u$ consisting of two
$2$-counters with the same value; more precisely a suffix of the form
$\s_3 v\s'_3 v \s''_3$ with $v$ a $2$-counter. If $u$ does not end
with two $2$-counters then \one\ can check this with $\counter_2$ test
and win.  If $u$ indeed ends with two $2$-counters then \one\ needs to
show that the values of these counters differ. For this he selects, by
removing letters from the store, a position in the topmost counter
where he thinks that the difference occurs. So the store now finishes
with $\s v\s'$, where $\s,\s'\in \S_2$ and $v$ is a $1$-counter. Next
\one\ performs $push_2$ operation which makes a ``copy'' of $1$-store.
The result is:
\begin{equation*}
  [u'\s v\s'][u'\s v\s']\ .
\end{equation*}
It is then the turn of \zero\ to pop letters from the copy of the
store in order to find in the second counter the position with number
$v$. We can be sure that \zero\ stops at some position of the second
counter by demanding that in the process he pops precisely one letter from $\S_3$.
After this the store has the form:
\begin{equation*}
  [u'\s v\s'][u''\r w\r']\ .
\end{equation*}
From this configuration \zero\ wins iff $v=w$ and $\s'=\r'$. This test
can be done in the same way as $\equal_1$ test.
\\[1 em]
Using similar techniques, it is also possible to define a test
$\Succ_2$ checking that the two topmost $2$-counters represent
successive numbers (from $[0,2^{2^n}-1]$).
%
%
%
\subsection{Counters of Higher Levels} \label{sec:kcount} 
%
%
%

As expected $k$-counters are defined by induction.
\debdef
For all $k>1$ a {\em $k$-counter} is a sequence of $(k-1)$-counters of the
form:
\debeqno
        \sigma_j \ell_j \cdots  \sigma_1 \ell_1 \sigma_0 \ell_0\ , 
\fineqno
where $j=Tower(k-1,n)-1$, for all $i\in [0,j] : \sigma_i \in \Sigma_k$ and
$\ell_i$ is a  
$(k-1)$-counter representing the number $i$.
This $k$-counter represents the number 
$\sum_{i=0}^{j}\sigma_i 2^i$. 
\findef
To cope with $k$-counters, $k$-HPDS are needed. We want to define for
all $k\pgd 2$ a $k$-HPDS with the control states with the following
properties:
\debitem
\item from $(\counter_k,u)$ \zero\ wins iff $u$ ends with a
  $k$-counter;
  
\item from $(\first_k,u)$, $(\last_k,u)$ \zero\ wins iff $u$ ends with
  a  $k$-counter representing $0$ and the maximal value
  respectively;
\item from $(\equal_k,u)$ 
  \zero\ wins iff the two last $k$-counters in $u$ have the same value;

\item from $(succ_k,u)$ \zero\ wins iff 
  the two topmost $k$-counters represent successive numbers.
\finitem

This is done by induction on $k$, using hypotheses for lower levels as
subprocedures. For $k=1$ and $k=2$, we have shown the constructions in
the previous subsection. In the following we consider some $k>2$ and
explain now the construction by induction.

Starting from a configuration $(\counter_k,u)$ we want \zero\ to win
iff on the top of the stack there is a $k$-counter; more precisely
that $u$ has a suffix $\s_{k+1} v \s'_{k+1}$ with
$\s_{k+1},\s'_{k+1}\in \S_{k+1}$ and $v$ a $k$-counter. A $k$-counter
is a sequence of $(k-1)$-counters, and the task of \one\ is to show
that $u$ has no suffix of the right form. One way to do this is to
show that $u$ does not end with a $(k-1)$-counter or that this last
counter does not have value $Tower(k-1,n)-1$. This \one\ can do with
$\last_{k-1}$ test.  Otherwise \one\ can decide to show that there is
some part inside the hypothetical $k$-counter that is not right. To do
this he is allowed to take letters from the stack up to some $\S_k$
letter at which point he can check that the two consecutive topmost
$(k-1)$-counters have wrong values (using test $\Succ_{k-1}$).  This
test can be performed only if \zero\ does not claim that the counter
on the top represents $0$.  If \zero\ claims this then \one\ can
verify by using test $\first_{k-1}$.  Similarly we can define
$\first_k$ and $\last_k$ test.

Next we want to describe $\equal_k$ test for which we will need
the power of $k$-stores. We want \zero\ to win from a configuration
$(\equal_k,u)$ iff there is a suffix of $u$ consisting of two
$k$-counters with the same value; more precisely a suffix of the form
$\xi z\xi' z \xi''$ with $z$ a $k$-counter and $\xi,\xi,\xi''\in
\S_{k+1}$. If $u$ does not end with two $k$-counters then \one\ can
check this with $\counter_k$ test and win.  If $u$ indeed ends with
two $k$-counters then \one\ needs to show that the values of these
counters differ. For this he selects, by removing letters from the
store, a position in the topmost counter where he thinks that the
difference occurs. So the store now finishes with $\s v\s'$, where
$\s,\s'\in \S_k$ and $v$ is a $(k-1)$-counter. Next \one\ performs
$push_2$ operation which makes a ``copy'' of $1$-store.  The result
is of the form:
\begin{equation*}
  [u'\xi z \xi' z' \s v \s'][u'\xi z \xi' z' \s v \s']\ .
\end{equation*}
This is a $2$-store with two elements where $z$ is a $k$-counter and
$z'$ is a prefix of a $k$-counter.

It is then the turn of \zero\ to pop letters from the copy of the
store in order to find in the second counter the position with number
$v$. We can be sure that \zero\ stops at some position of the second
counter by demanding that in the process he pops precisely one letter
from $\S_{k+1}$.  After this the store has the form:
\begin{equation*}
  [u'\xi z \xi' z' \s v \s'][u'\xi z''\r w \r']\ .
\end{equation*}
From this configuration \one\ wins iff $v\not=w$ or $\s'\not=\r'$.
Checking $\s'\not=\r'$ is easy. The test whether $v=w$ can be done in
a similar way as $\equal_{k-1}$ test. The difference is that now we
have $2$-store and $\equal_{k-1}$ works on $1$-stores. We elaborate
the construction as this is the place where the power of $k$-stores
really comes into play.

We will construct states $same^i_k$, for $i< k$, with the property that \zero\ wins
in a configuration with a $(k-i+1)$-store of the form
\begin{equation*}
  s[u\struct{r\s v\s'}][u'\struct{r'\r w\r'}]\ .
\end{equation*}
iff $\s'=\r'$ and $v=w$ is a $i$-counter. Here $\s,\s',\r,\r'\in
\S_{i+1}$, $r$, $r'$ are sequences of letters, $u$, $u'$ are
$(k-i)$-stores and $s$ is a $(k-i+1)$-store.  The notation $\struct{\s
  v\s'}$ is to denote the first 1-store in the given store, hence
$\struct{\ }$ stand for some number of nested $[\ ]$ parentheses. The
verification we need in the last paragraph is precisely $same^{k-1}_k$
as there we have a $2$-store and compare $(k-1)$-counters.

It is quite straightforward to construct $same^1_k$. \one\ has the
right to declare that either $\s'\not=\r'$ or that the counters are
not equal. Checking the first case is straightforward. To show that
the counters are different, \one\ chooses $j\leq n$ and pops $j$
letters from $w$ using $pop_1$. Then $j$ and the top letter are
remembered in the control state.  Afterward $pop_{k-1}$ 
is performed and once more $j$ letters
are popped.  \one\ wins if the top letter is different from the one
stored in the finite control.

To construct $same^{i}_k$ for $i>1$ we proceed as follows. \one\ has
the possibility to check if $\s'=\r'$ as before. The other possibility
is that he can $pop_1$ some number of letters finishing on a letter
from $\S_{i}$ and without popping a letter from $\S_{i+1}$ in the
process. The resulting configuration is of the form:
\begin{equation*}
  s[u\struct{r\s v\s'}][u' \struct{r'\r w'\t x\t'}]\ .
\end{equation*}
The intuition is that \one\ declares that at position $x$ in $v$ the
value is different than $\t'$. Now $push_{k-i+2}$ is performed
giving the configuration
\begin{equation*} 
  \big[s[u\struct{r\s v\s'}][u' \struct{r'\r w'\t x\t'}]\big]\ %
  \big[s[u\struct{r\s v\s'}][u'\struct{r'\r w'\t x\t'}]\big] \ .
\end{equation*}
As we had $(k-i+1)$-store before, now we have $(k-i+2)$-store
consisting of two elements.  

Next we let \zero\ to do $pop_{k-i}$ and some number of $pop_1$
operations to get to the situation
\begin{equation*}
  \big[s[u\struct{r\s v\s'}][u' \struct{r'\r w'\t x\t'}]\big]\ 
  \big[s[u\struct{r\s v'\g y\g'}]\big]\ .
\end{equation*}
where he claims that $x=y$ and $\t'=\g'$. This can be checked from
$same^{i-1}_k$ state.

The procedure $succ_k$ is implemented similarly to $equal_k$. Here it
is not the case that at each position in the counters bits should
be the same.  Nevertheless the rule for deducing which bit it should
be is easy and the difficult part of comparing the positions is done
using $same^{k-1}_k$.

%
%
%
\section{Encoding Turing Machines} \label{sec:TM} 
%
%
%

In this section we will show how to encode computations of an
\EXPSPACE-bounded Turing machine using $2$-store. Then we will claim
that the construction generalizes to alternating $k$-\EXPSPACE\ and
$(k+1)$-stores.

Fix $M$, an \EXPSPACE-bounded alternating Turing machine (TM), as well as
an input word of length $n$.  The set of control states of the TM is
denoted $Q$. A {\em configuration} of $M$ is a word over
$\D_2=\{a_2,b_2\}\cup Q \cup\{\vdash,\dashv\}$ of the form 
\debeqno 
      \vdash u_1 \cdots u_i q u_{i+1}\cdots u_j \dashv 
\fineqno 
where $q\in Q$,
$\forall k: u_k\in \{a_2,b_2\}$. Here the TM is in state $q$, reading
letter $u_{i+1}$.  

We will encode configurations of $M$ almost in the form
of $2$-counters to write them in the store of a HPDS. Let $k=2^n$. 
A configuration $\sigma_0\sigma_1\cdots \sigma_{k-1}\in (\D_2)^k$ 
is represented by a word
\debeqno
        \xi \sigma_{k-1} \ell_{k-1} \cdots  \sigma_1 \ell_1 \sigma_0 \ell_0\xi\ ,
\fineqno
where for all $i\in [0,2^n-1]$:
$\sigma_i \in \D_2$, $\ell_i\in (\Sigma_1)^n$ is a 
$1$-counter representing the number $i$, and 
$\xi\in \Sigma_3$ is a separator.

A computation is represented as a string obtained by concatenation of
configurations.  The game will proceed as follows: departing from the
initial configuration of the Turing machine (the input word), Player~0
is in charge of building an accepting run and Player~1 is in charge of
checking that no error occurs. \zero\ simply writes letter by letter a
configuration. If the state of the configuration is existential then
after writing down the configuration \zero\ writes also a transition
he wants to perform. Otherwise it is \one\ who writes the transition.
Then \zero\ continues with writing a next configuration that he claims
is the configuration obtained by the transition that was just written
down. This process continues until a configuration with a final state
is reached. At the end of writing each configuration \one\ has the
opportunity to check if the last two configurations on the stack
indeed follow from each other by the transition that is written
between them.

Let us describe some details of this construction. Applying a
transition rule of the Turing Machine consists in rewriting only three
letters: $u_i$, $q$ and $u_{i+1}$ in the notation of the example
above. To check that the transition is legal, we will proceed in
several steps. After writing a configuration, ended by a separator
$\xi\in \S_3$, \zero\ has to write again the three letters $u_i q
u_{i+1}$. Then, depending whether state $q$ is existential or
universal in the TM, \zero\ or \one\ writes three other letters of
$\D_2$, say $q'a c$, such that $(u_i q u_{i+1},q'ac)$ is a transition
rule of the TM. The other player can test that this transition rule is
indeed in the TM.

After that \zero\ has to write down the configuration obtained by the
chosen transition, and \one\ has the opportunity to test whether this is
correct. To do this he has several possibilities.  First he can check
that the newly written configuration is of a correct form, using a test
similar to $counter_2$, replacing $\S_2$ by $\D_2$.

Otherwise he can check that this two last configurations are identical,
except for the part involved in the transition rule. The store at
this point is:
\begin{equation*}
  s\ \xi c_1 \xi\  u_i q u_{i+1}q'ac\  \xi  c_2  \xi\ ,
\end{equation*}
where $s$ is a prefix of computation, $c_1$ and $c_2$ are the last two
configurations separated by the chosen transition. We describe a game
from a state $trans_2$ such that \zero\ wins from $trans_2$ and the
store as above iff the two topmost configurations obey the transition
rule written between them. The test $trans_2$ has the same structure
as the test $equal_2$. \one\ has first to pop letters to select a
position in the configuration, that is a 1-counter. Each time he wants
to pop next 1-counter he asks \zero\ if this position is the
rightmost position involved in the transition or not. If yes then
\one\ has to pop three counters at the time, if not he pops one
counter. Finally, \one\ stops at a position where he thinks that an
error occurs. He asks \zero\ if this position is the rightmost
position of the transition. If \zero\ says that it is not then it is
tested that at the same position in the preceding configuration there
is the same letter; this is done in the same way as $equal_2$ test.

If \zero\ claims that the chosen position is the rightmost position of
the ones involved in the transition then the test is slightly more
complex. A $push_2$ is performed and the store becomes
\begin{equation*}
  [s\ \xi c_1 \xi\ u_i q u_{i+1}q'ac\ \xi c_2'\ \r v \r' v' \r'' v'' ] 
  [s\ \xi c_1 \xi\ u_i q u_{i+1}q'ac\ \xi c_2'\ \r v \r' v' \r'' v'' ]\ ,
\end{equation*}
where $c_2'$ is a prefix of $c_2$, $\r,\r',\r'' \in \D_2$ and
$v,v',v''$ are $1$-counters. \one\ has the opportunity to check that
$q'ac=\r\r'\r''$, which is easy to implement. \one\ has also the
opportunity to let \zero\ find the position in $c_1$ corresponding to
$v''$ and then test that the corresponding letters from $\D_2$ are
exactly $u_i q u_{i+1}$; this is implemented in a similar way as in
$equal_2$ test.

The game is won by \zero\ iff he can write an accepting configuration
of the TM without \one\ ever challenging him, or if \one\ fails in some
test.  In other words the game is won by \one\ iff he can prove that
\zero\ was cheating somewhere or if \zero\ never reaches an accepting
configuration of the TM. Examining the construction one can see that
we need $\Oo(n^2+|M|)$ states in $2$-HPDS to carry out the described
constructions. So we have a poly-time reduction of the acceptance problem of
alternating \EXPSPACE{} Turing Machines to the problem of determining
the winner in a reachability game over a $2$-HPDS.
\debthm
Reachability games on $2$-HPDS are $2$-\EXPTIME\ hard.
\finthm
Together with the double exponential time solution of the more 
general parity games from \cite{icalp03}, we have:
\debco
Reachability/parity games on $2$-HPDS are complete for $2$-\EXPTIME.
\finco
Using the constructions of Section~\ref{sec:kcount}, it is easy to
extend the encoding above and show that alternating $k$-\EXPSPACE\ 
Turing Machines can be simulated by $(k+1)$-HPDS. Together with the
results from \cite{icalp03} we get:
\debthm
Reachability/parity games on $k$-HPDS are complete for $k$-\EXPTIME.
\finthm
This result gives also a new proof that the hierarchy of HPDA is strict, and
together with \cite{icalp03}, that the Caucal hierarchy is also strict.

%
%
%
\section{Conclusion} 
%
%
%

The $k$-\EXPTIME\ lower bound that we have proved in this paper shows that
games are difficult on HPDA, even the simplest ones : reachability games.
Surprisingly the complexity for solving parity games is the same as for
reachability games. It is open to find algorithms or lower bounds for the model
checking of other logics like CTL or LTL, that are weaker than the
$\m$-calculus.

%
%
%
\subsection*{Acknowledgment}
%
%
%

Many thanks to Luke Ong and Olivier Serre for interesting discussions.

%
%
%

\end{document}